# Open source QGIS toolkit for the Advanced Research WRF modelling system

## To cite this article

## BibTeX

```
@article{Meyer2019,
  doi = {10.1016/j.envsoft.2018.10.018},
  issn = {1364-8152},
  url = {https://doi.org/10.1016/j.envsoft.2018.10.018},
  year  = {2019},
  month = {feb},
  publisher = {Elsevier {BV}},
  volume = {112},
  pages = {166--178},
  author = {D. Meyer and M. Riechert},
  title = {Open source {QGIS} toolkit for the Advanced Research {WRF} modelling system},
  journal = {Environmental Modelling {\&} Software}
}
```

## Copyright and License

# Open source QGIS toolkit for the Advanced Research WRF modelling system

D. Meyer[a] and  M. Riechert[b]

[a] Department of Meteorology, University of Reading, Reading, UK

[b] Microsoft Research, Cambridge, UK

## Abstract

The Advanced Research WRF (Weather Research and Forecasting) model is a popular atmospheric model used for research and Numerical Weather Prediction (NWP). However, despite its popularity, its set-up and configuration often demand several interdisciplinary skills that go beyond the understanding of physical processes. Pre-processing tasks, such as importing custom high-resolution datasets in the WRF Pre-processing System (WPS), still require a considerable effort from the user. We present GIS4WRF, a free, open-source, and cross-platform QGIS Python plug-in to help scientists and practitioners with their Advanced Research WRF modelling workflows. GIS4WRF incorporates new and existing tools for data-processing, configuration, simulation, and visualization into a single graphical environment, and offers WRF-CMake binary distributions for Windows, macOS, and Linux. We highlight its main features and provide useful insights into several key approaches and techniques used in its development. We end with two example applications highlighting the contributions of GIS4WRF in simplifying several WRF-related tasks.

# 1 Introduction

The Weather Research and Forecasting model (WRF) (Janjić et al., 2010; Skamarock et al., 2008) is a popular, free and open source, community-driven atmospheric modelling system aimed at research and numerical weather prediction (NWP) applications. First released in 2000, it is the result of more than two decades of development, vaunting a role in over 3,500 peer-reviewed publications from over 11,700 unique authors worldwide (Powers et al., 2017).

WRF has been developed under two variants, found in its two fluid-flow solvers, or dynamical cores: the Advanced Research WRF (ARW) core (Skamarock et al., 2008, 2005), and the Non-hydrostatic Mesoscale Model (NMM) core (Janjić, 2003; Janjić et al., 2001) whose support has recently ended (Developmental Testbed Center (DTC), 2018). Since its release, its capabilities have been greatly enhanced by the addition of several components such as the WRF Data Assimilation System (WRFDA; Barker et al., 2012, 2004; Huang et al., 2009), the WRF-based atmospheric chemistry model WRF-Chem (Fast et al., 2006; Grell et al., 2005), the WRF Hydrological Modelling Extension Package (WRF-Hydro; Gochis et al., 2015), WRF-Fire (Coen et al., 2013), Hurricane WRF (HWRF; Biswas et al., 2017), WRF-Urban (Chen et al., 2011), and WRF-Solar (Jimenez et al., 2016). Some of these components, such as WRF-Urban, are released and bundled together with the official software distribution of ARW or NMM, while others, such as HWRF, are maintained and released as separate packages. Here, we exclusively discuss the official distribution of ARW, and simply refer to it as WRF.

As a flexible model, WRF can be configured to conduct idealized and real-data simulations. Idealized configurations provide users with configurable user-generated initial conditions to study processes in simplified situations, i.e. lab-type experiments. Real-data simulations allow users to use initial conditions assimilated from observations, and land surface characteristics derived from geographical datasets, thus representing real-world settings. Where idealized simulations are configured from human-readable text files in WRF, real-data cases require several pre-processing steps to ingest meteorological and geographical data before the simulation can begin. In this work, we will mainly focus on real-data workflows.

Typical real-data workflows involving the use of WRF are characterized by three main steps (Fig. 1): pre-processing, simulation, and post-processing. During pre-processing, users define the characteristics of their computational domains together with their choice of data in a configuration file (*namelist.wps*), then, in the WRF Preprocessing System (WPS), the model's computational domains are created and the geographical data (e.g. topography and land use) are mapped by the *geogrid* program. Atmospheric data (e.g. global analyses or forecasts) are first converted by the *ungrib* program, and horizontally interpolated to the model domain by the *metgrid* program. Next, the pre-processed files, together with a user-defined configuration file (*namelist.input*) are used to vertically interpolate data and create lateral boundary conditions using the *real* program. Finally, the *wrf program is* configured using the *namelist.input* file with additional information about physical parametrizations (e.g. microphysics, radiation, planetary boundary layer), and run to produce the model forecast, which can be later visualized or further post-processed with external tools, e.g. to calculate wind speed and direction.

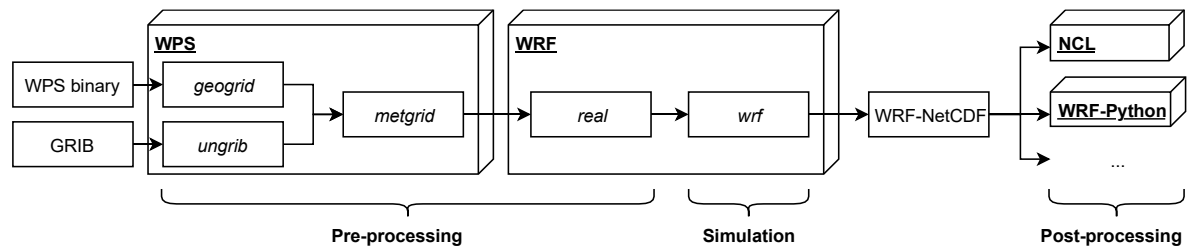

*Fig. 1. Typical WRF workflow and artefacts*

In WPS and WRF workflows, files are typically handled using a common format except for two WPS programs: *geogrid* and *ungrib*. Where *ungrib* accepts atmospheric data in the widespread GRIB format (World Meteorological Organization, 2003), *geogrid* can only read data in a bespoke format, named WPS binary, that represents gridded two- or three-dimensional data as multiple binary files and a metadata file (Wang et al., 2017). For all other programs, input and output data are stored in the NetCDF format (Unidata, 2017) using a custom metadata convention, here named WRF-NetCDF. This convention differs from the well-defined and widely used Climate and Forecast (CF) Metadata Conventions (Eaton et al., 2017) in the way metadata for some variables are represented. For example, WRF-NetCDF files store the Coordinate Reference System (CRS) definition as a set of global attributes, whereas the CF Metadata Conventions require that the CRS is represented in a so-called *grid mapping* variable.

Although the configuration of namelists in WPS and WRF workflows merely requires users to follow the documentation, other tasks, such as the building of executables, the configuration of domains, the pre-processing of geographical data, and the visualization or post-processing of outputs may be viewed as non-trivial and time-consuming tasks that often require users to be familiar with several interdisciplinary skills, beyond a good understanding of physical processes. As an example, the ability to use bespoke geographical datasets in weather modelling has become essential in initializing realistic simulations as the spatial resolution of models increases (Beezley et al., 2011; Mandel et al., 2011), however, this requires conversion of existing datasets to the WPS binary format and an understanding of technical details such as tiling, halo width, scaling factor, and coordinate conversion.

To facilitate the use of WPS and WRF, tools have been developed to help with the workflow and scheduling of jobs, e.g. WRF Portal (Govett and Smith, 2007; Smith et al., 2007); the configuration of domains and namelists, e.g. WRF Domain Wizard (Smith et al., 2007); visualization, e.g. WRF Domain Wizard, NCL (Brown et al., 2017), WRF-Python (Ladwig, 2017), Vapor (Clyne et al., 2007; Clyne and Rast, 2005); and post-processing of outputs, e.g. NCL, WRF-Python and Vapor. These tools however are generally confined to a specific purpose, whether this is pre- or post-processing, or job scheduling. For example, WRF Domain Wizard and WRF Portal can be used to configure namelists, run simulations, and visualize results, yet they lack support for importing and exporting geographical datasets. NCL, Vapor and WRF-Python can visualize WRF-NetCDF files, but their use is limited to visualization and post-processing. Indeed, none of these tools let users import, export, convert, or visualize WPS binary format datasets. To our knowledge, the only two relevant developments to address the import of third-party, including high-resolution, georeferenced data have been made by Beezley (2013; 2011). He modified the *geogrid* program to allow users to import a GeoTIFF dataset directly into WRF and released a command-line utility written in C to convert GeoTIFF to WPS binary that supports high-resolution datasets using the tiling feature of WPS binary. What these tools lack, however, is the integration with the official WPS source code or other tools. For example, they need to be compiled manually and cannot be used to visualize existing or converted WPS binary datasets.

In our opinion, a toolkit that incorporates new and existing WRF tools for data processing, configuration, simulation, and visualization into a single graphical environment is key in making WRF more accessible whilst empowering users, who may lack the sufficient technical knowledge, to carry out more advanced tasks such as using bespoke, or high-resolution data in their simulations.

In our initial assessment, we considered extending WRF Portal and WRF Domain Wizard by adding support for georeferenced data together with a more detailed representation of maps. We found, however, that the integration with other useful tools such as WRF-Python would have been difficult. Moreover, the choice of Java, in the case of WRF Portal and WRF Domain Wizard, as opposed to other, more commonly-used languages in the scientific community, such as Python, might have deterred users from actively contributing to the development of the software. For these reasons, we concluded that, although WRF Portal and WRF Domain Wizard were already well-developed, the downsides of having a software difficult to maintain, or to extend in the future, did not make WRF Portal and WRF Domain Wizard a viable option in our opinion. We did not consider the prospect of extending NCL or WRF-Python as their purpose is exclusively on visualization and post-processing.

In recent years, the potential for adding functionalities to GIS (Geographical Information System) software for environmental modelling applications has been highlighted by several tools such as QSWAT (Dile et al., 2016), UMEP (Lindberg et al., 2018), WET (Nielsen et al., 2017), QMorphoStream (Tebano et al., 2017) and STAR-BME (Yu et al., 2016). A review by Chen et al. (2010), has positively found QGIS (QGIS Development Team, 2018a) to be a good alternative to popular commercial GIS software such as ArcGIS (Environmental Systems Research Institute (ESRI), 2018). Although their review focused on water resources management in developing countries, several features they tested are relevant and common to several scientific areas. QGIS is a cross-platform, free and open source desktop GIS application written in C++ and Python used by a wide variety of users on Windows, macOS, and Linux systems. The functionalities of QGIS are greatly increased by the use of its Python API (Application Programming Interface) for creating plug-ins. Since their introduction in 2007, (Graser and Olaya, 2015) the number of QGIS plug-ins have increased rapidly. Counting more than 950 at the time of publication (QGIS Development Team, 2018b), QGIS plug-ins offer users a vast number of added capabilities within a single framework.

Here, we present GIS4WRF, a free, open source, and cross-platform toolkit developed as a QGIS Python plug-in to help scientists and practitioners with their WRF workflows in pre- and post-processing data, simplifying the simulation steps and visualizing and post-processing their model results. GIS4WRF aims to solve the above-mentioned issues found in other software by bringing together new and existing WRF tools into a single graphical environment in QGIS. By relying on the QGIS framework, GIS4WRF lets users benefit from QGIS's rich raster support, and many other features like re-projection, resampling, on-the-fly visualization, merging and comparison of datasets, and integration with existing plug-ins.

The paper continues with a general overview of features included in GIS4WRF, and information on how to install and get started with the plug-in in section 2. From this general overview, we follow with a more detailed description of novel features in section 3. In section 4 we give an insight into our software development process and discuss certain aspects of the implementation; users who wish to develop or improve QGIS plug-ins may find this content of value. We conclude with two example applications showcasing how GIS4WRF can be used in section 5 and ideas for future development in section 6.

## 2 General Overview and Installation

GIS4WRF is a QGIS plug-in written in Python aimed at supporting researchers and practitioners in preparing input data for WRF simulations, running WRF simulations, and viewing, analyzing, and post-processing results from WRF simulations through a single user interface in QGIS. GIS4WRF is built as a free, open source, and cross-platform QGIS plug-in, allowing for easy installation, automatic update notifications, and use of built-in QGIS functionality and access to the QGIS plug-in ecosystem. After installing QGIS, GIS4WRF can be downloaded and installed directly from the QGIS plugins manager found under the *Plugins > Manage and Install Plugins …* menu.

GIS4WRF includes several features organized in a detachable, resizable, and movable container in QGIS, referred to as dock. The GIS4WRF dock is composed of four main tabs: *Home*, *Datasets*, *Simulation*, and *View*. Apart from the *Home* tab, which is designed to welcome users and provide support and useful documentation, all features in GIS4WRF can be grouped into three categories: (A) Datasets Preparation, (B) Simulation Setup, and (C) Visualization. These feature-categories are represented as such in the graphical user interface (GUI) (Fig. 2), and follow a common workflow used by researchers when preparing, running, and visualizing results from simulations, and should therefore feel intuitive to the user. Table 1 provides an overview of all features currently available and their corresponding location in the GIS4WRF dock.

*Table 1*
*Features currently available in GIS4WRF. Features are grouped into three main categories and are represented as such in the main GIS4WRF graphical user interface as tabs or subtabs (Fig. 2).*

| ***Category (Tab)**/Feature* |
|---|
| **A. Datasets Preparation (Datasets)** |
|     1. Download WPS geographical datasets |
|     2. Download meteorological datasets[1] |
|     3. Convert raster data (e.g. GeoTIFF) into WPS geographical datasets |
| **B. Simulation Setup (Simulation)** |
|     1. Load/save simulation setup from/to GIS4WRF project file |
|     2. Define domains |
|     3. Visualize domains as vector layers |
|     4. Import/export domains from/to *namelist.wps* files |
|     5. Define geographical and meteorological datasets |
|     6. Prepopulate configuration files (*namelist.wps*, *—.input*) for running WPS and WRF |
|     7. Download pre-built WPS/WRF binaries for Windows, macOS and Linux |
|     8. Run WPS and WRF (from pre-built or existing installations) |
| **C. Visualization (View)** |
|     1. Open WRF-NetCDF files as multiple raster layers |
|     2. Control the visualization of variables and dimensions in WRF-NetCDF files |
|     3. Compute diagnostics with WRF-Python |
|     4. Open WPS geographical datasets as raster layers |

---

[1] Currently, this feature only supports the download of specific datasets from the National Center for Atmospheric Research (NCAR) Research Data Archive (RDA; https://rda.ucar.edu).

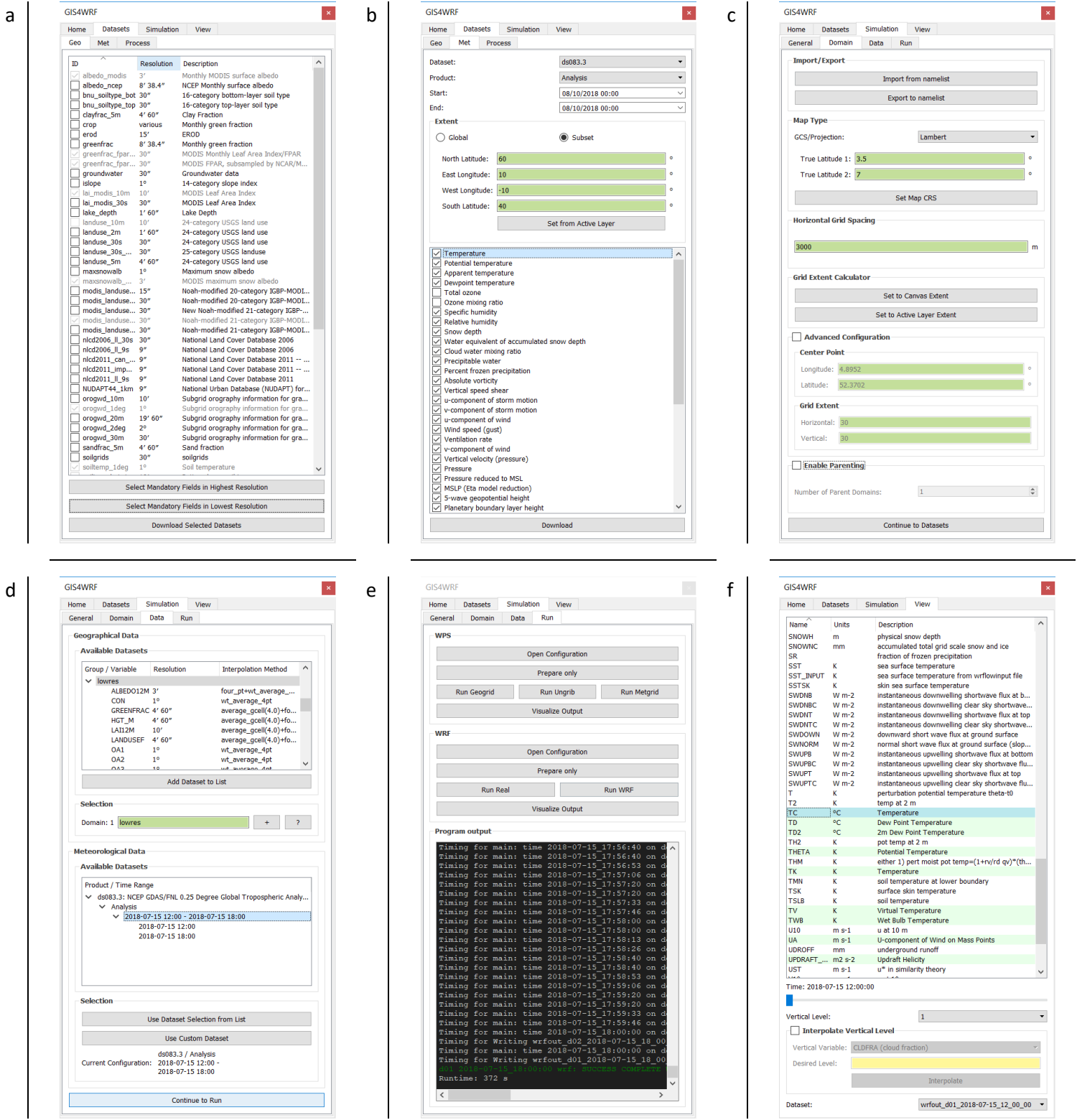

*Fig. 2. Overview of the main components described in Table 1 as implemented in GIS4WRF. Each of the following tab/subtabs pictured above contains one or more features in Table 1. (a) Geo subtab to download WPS geographical datasets; (b) Met subtab to download meteorological datasets; (c) Domain subtab to load/save simulation setup from/to GIS4WRF project file, define domains, visualize domains as vector layers, and import/export domains from/to namelist.wps files; (d) Data subtab to define geographical and meteorological datasets; (e) Run subtab to prepopulate configuration files (namelist.wps, —.input) for running WPS and WRF, and run WPS and WRF (from pre-built or existing installations); (f) View tab to control the visualization of variables and dimensions in WRF-NetCDF files, and compute diagnostics with WRF-Python. The Home tab, Process subtab, and General subtab are not shown here.*

The installation of GIS4WRF is made effortless by the full integration with the QGIS ecosystem. Users are only required to install the latest version of QGIS 3 (see https://download.qgis.org) on their systems. GIS4WRF can be installed directly from the *Manage and Install Plugins* window under the *Plugins* menu in QGIS. During the installation process, GIS4WRF automatically checks the system and downloads any required dependencies (see the Software Development section for an in-depth explanation). Once the installation is successful, users can launch GIS4WRF from its icon button (🌐) in the *Plugins Toolbar* or from the menu in *Plugins > GIS4WRF.*

# 3    Description of Features

The main features of GIS4WRF allow for easy conversion and manipulation of spatial datasets, definition and visualization of domains, generation of name-lists, and native integration with the QGIS framework for displaying geospatially located WRF input and output data as raster layers. Furthermore, users can optionally download the pre-built WPS and WRF binaries for Windows, macOS, or Linux generated with WRF-CMake (see https://github.com/WRF-CMake), or link to their own compiled version of WPS or WRF to carry out the simulation steps in GIS4WRF. In this section, we present and discuss features in each category as summarized in Table 1. An exhaustive documentation of features and GUI elements is outside the scope of this paper and can be found in the online manual (see https://gis4wrf.github.io/documentation).

## 3.1    Datasets Preparation

Tools to download geographical and meteorological datasets, and convert to WPS binary format are organized in the *Datasets* tab. This tab is further split into three subtabs: *Geo*, *Met*, and *Process*. We support the download of WPS geographical input data from the *Geo* subtab, the download of several meteorological datasets from the *Met* subtab, and the conversion of QGIS raster layers to WPS binary datasets from the *Process* subtab.

From the *Geo* subtab (Fig. 2a), users can download official WPS geographical input datasets supported by the WRF Users Page (http://www2.mmm.ucar.edu/wrf/users). Downloads are automatically saved under the user-defined working directory and made available to the user through the GIS4WRF interface. Users can optionally double-click on the downloaded dataset to load the files as a QGIS raster layer (see Visualization section).

GIS4WRF includes integration with the Research Data Archive (RDA; https://rda.ucar.edu) download client API through the interface in the *Met* subtab (Fig. 2b). The RDA provides users with a portal to find and download diverse meteorological datasets. For unrestricted datasets, the service is free of charge, however, users must register before they can start downloading data or use the service in GIS4WRF. Users can use the *Met* subtab to download currently supported[2] datasets (e.g. NCEP (National Centers for Environmental Prediction) Analysis) to provide initial and boundary conditions for their WRF simulations. The interface aims at simplifying the process of downloading data within a single GUI. In addition, we have added support to subset download requests by using a QGIS layer; the feature is fully integrated with the domain tools (see Simulation Setup section) and allows users to simplify and minimize the size of their download requests. Users wishing to use other datasets, not available through the RDA integration (e.g. ECMWF datasets), can still use GIS4WRF and import datasets manually (see *Data* in Simulation Setup).

The *Process* subtab contains a tool for converting QGIS raster layers to WPS binary datasets. QGIS raster layers are created by opening geospatial datasets using the Geospatial Data Abstraction Library (GDAL) (Open Source Geospatial Foundation, 2018) and, as such, any of the formats supported by GDAL[3], e.g. GeoTIFF, are also supported in QGIS. We support both continuous (e.g. elevation) and categorical (e.g. land use) data. The integration with GIS4WRF visualization tools (see Visualization) allows users to view their converted datasets before using them in their simulation. The tool aims at simplifying the conversion of data, however, it is not meant to be a replacement for scientific judgment. Users are strongly advised to review the possible pitfalls when dealing with data from different sources that may have been produced with different datum or CRS assumptions than the default datasets used in WPS (see Monaghan et al. (2013)).

### 3.2 Simulation Setup

The simulation tab in GIS4WRF contains tools to help users prepare their simulations. It is formed of four other subtabs, namely *General*, *Domain*, *Data*, and *Run*. The *General* subtab is reserved to creating and opening GIS4WRF projects, whereas *Domain*, *Data* and *Run* contain several tools to help users with all the required steps to define and run simulations in WRF.

**Domain**

The *Domain* subtab (Fig. 2c; Fig. 3 – right) includes features aimed at helping users with creating, exporting, and importing WRF domains. Our principle motivation is to simplify the creation of WRF domains and to integrate support and interaction with other QGIS and GIS4WRF tools. For example, users can display domains as vector layers choosing any WRF-supported projections, configure single and nested domains interactively using a GUI, preview changes to their domain configurations instantly on the map canvas, and import or export domain set-ups from or to WPS namelists.

---

[2] For an up-to-date list of supported datasets, please refer to the online documentation at https://gis4wrf.github.io/documentation.
[3] See https://www.gdal.org/formats_list.html for a full list of supported formats.

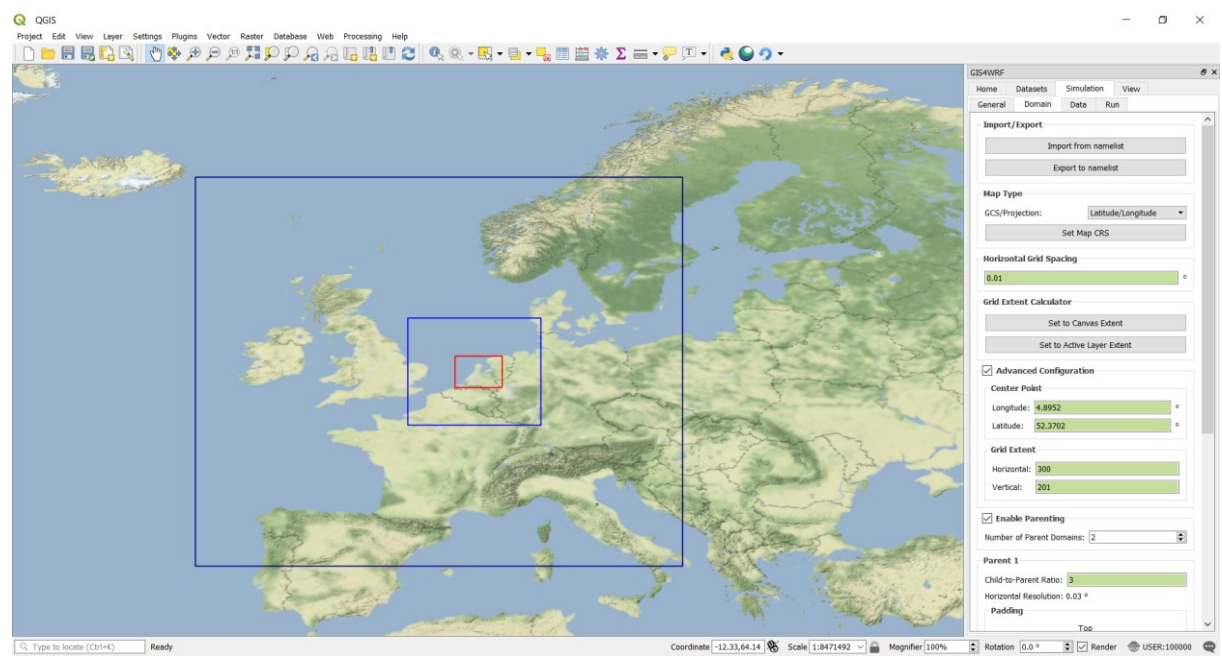

*Fig. 3. GIS4WRF Domain subtab (right panel), QGIS canvas with Stamen base map (Stamen Design, 2018) and 3 domains (center), and QGIS's Layers panel (left).*

Each WRF domain is drawn as a separate vector layer in QGIS (Fig. 3 – center) while user inputs are automatically validated and, if necessary, adjusted to conform with WRF domain requirements. Users can use WRF domain layers as part of their workflow in QGIS or export them to other software. For example, the GIS4WRF meteorological data download tool (see Datasets Preparation section) already integrates with this feature allowing users to limit their data request to the extent defined by their domains.

We tried to keep the workflow as simple as possible while allowing for full customization. Users can get started with new domains using the *Set to Canvas Extent* button and further refine them under *Advanced Configuration*. This option allows novice and expert users to get started rapidly with an overview of their region of interest and consider how characteristics, such as land surface, can affect their choice of extent and location for a given computational domain. Users can create domains based on the resolution and extent of other layers by using the 'Set to Active Layer' button. Furthermore, users who wish to load an existing *namelist.wps* file in GIS4WRF, or save their domain configuration as *namelist.wps* file, can make use of the importer and exporter buttons.

An important change introduced in GIS4WRF is the way nested-type WRF domains are defined. Differently to WRF, where nested-type WRF domains are defined based on the parent or outermost domain (PD or OMD), domains in GIS4WRF are always defined based on the child or innermost domain (CD or IMD). We believe that this change provides greater control with the position and extent of the IMD, i.e. the domain covering the area of interest. When standard WRF domain namelists are imported, GIS4WRF executes an implicit conversion between the WRF PD-based reference system and the GIS4WRF CD-based system. For calculations to move from one reference system to the other see Appendix – Domain Calculations.

When creating WRF domains, the *Domain* GUI requires users to choose a coordinate reference system supported by WRF, the horizontal grid spacing, the center point longitude and latitude, and the grid extent. Any further PDs are created based on the IMD together with the *child to parent ratio* and four *padding* values. The padding defines the number of grid cells (extent) of the PD at its horizontal grid spacing in the north, east, south, west direction from the respective edges of its CD. In GIS4WRF the padding is used in combination with the IMD center point, to calculate the position and extent of outer domains. With this approach, when changing the horizontal grid spacing or extent of a domain, the IMD remains centered as defined by the user, whereas in the OMD approach several values need to be adjusted to re-center the IMD to a specific location.

**Data**

The *Data* subtab (Fig. 2d) features several controls for setting the type of data to use in WPS and WRF. Users can display the data to use in their simulations that was previously downloaded, or manually converted, from the *Datasets* subtab. There are two main sections users have access to: the *Geographical Data* section, and the *Meteorological Data* section. Each of these sections can be completed independently and are aimed at helping users to pre-populate WPS and WRF namelists (*namelist.wps and namelist.input*), and *GEOGRID.TBL* files used to define fields produced by *geogrid*. For each domain generated in the *Domains* subtab, under the *Geographical Data* section, users can select the geographical dataset to use. Similarly, under the *Meteorological Data* section, users can view and select meteorological datasets previously downloaded from the Met subtab, or manually import other datasets, not currently available for download (e.g. ECMWF datasets).

**Run**

A key feature in GIS4WRF is the integration and packaging of pre-built WPS and WRF binaries for Windows, macOS and Linux through WRF-CMake (see https://github.com/WRF-CMake). Users are not required to use GIS4WRF pre-built binaries and can, instead, link to their own pre-built binaries. The *Run* subtab (Fig. 2e) allows users to specify the final configuration before running the executables.

Depending on the type of chosen configuration, it may be unfeasible to run a simulation on a local machine, in which case we suggest running all the preprocessing steps locally and continuing with the final simulation on a larger system. For example, users wishing to run simulations requiring many CPU-hours (e.g. ensembles, climate simulations) can first generate namelists and WRF input data by running WPS and *real* locally using GIS4WRF, and later transfer the content of the *run_wrf* folder – a folder containing WRF configuration and input/output files – to a cluster/super-computer where the most CPU-intensive work is executed. This allows users to optimize their time in making sure the configuration has been correctly defined and only carry out the simulation step on large systems, thus optimizing shared resources.

### 3.3 Visualization

GIS4WRF provides support for displaying georeferenced WPS binary and WRF-NetCDF files. Data can be imported from the *Layer > Add Layer* menu in QGIS while variables, timesteps and vertical-levels can be controlled from the *View* tab (Fig. 2f). All visualizations are computed in-memory without intermediate files, thus allowing for very fast opening times. All files are automatically georeferenced according to the information contained in the file metadata.

WPS binary files are rendered as raster layers and, for categorical data only, pre-defined legends and color maps are automatically generated for standard WPS land-use datasets (Fig. 4). QGIS efficiently renders raster layers of any size by loading only those subsets of data that are needed given the map canvas extent and zoom level. As WPS binary files are stored uncompressed, the rendering performance is typically limited by how fast data can be read from disk and how much memory is available for caching this data.

WRF-NetCDF files are displayed in the QGIS canvas as a raster layer. When WRF-NetCDF files are loaded, controls become available from the View tab. For example, users can slice through different timestamps using the time and date slider. Users have full control of more advanced display options such as color scales and projections through the native QGIS interface. As GIS4WRF comes already bundled with WRF-Python, we currently support a limited set of functions to compute diagnostics such as wind speed and direction, and pressure at different levels. Similar to WPS binary files, QGIS provides for efficient rendering by only loading those subsets of data that are needed, and, since WRF-NetCDF datasets are stored compressed the rendering performance is typically limited by the speed of the system's processor rather than the disk.

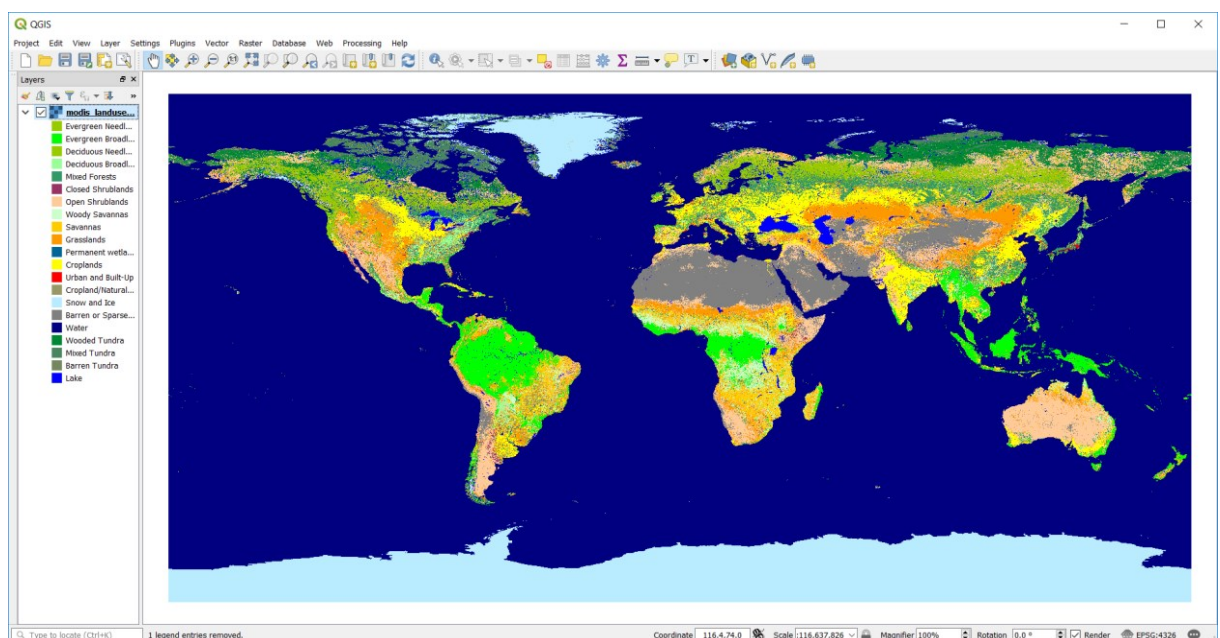

*Fig. 4. Visualization of WPS binary 20-class MODIS land use dataset in QGIS*

## 4   Software Development

A QGIS plug-in is a way to extend the base functionality of QGIS using its Python API. The complexity of plug-ins can vary greatly depending on the type of application: from adding a single button that carries a single purpose, e.g. copying the canvas into the clipboard (Thomsen and Municipality of Frederikssund, 2017), to rich toolboxes that provide a multitude of new functions, often exposed through rich GUIs, e.g. the Trends.Earth toolbox (Conservation International, 2018) for monitoring land change and preparing reports for government agencies.

GIS4WRF can be considered a complex plug-in given its number of features and rich, dynamically generated GUI. During the development of GIS4WRF we noticed a general lack of guidance and best practices for creating plug-ins of this complexity. Understandingly, given that most plug-ins are rather small in their structure and rely on unmodified templates created by the *Plugin Builder* plug-in in QGIS (GeoApt LLC, 2018), it is perhaps not surprising. However, we found that academic publications involving larger plug-ins (see Pumo et al. (2017), Sanzana et al. (2017)) published in respected software-related journals typically focus solely on their features, while missing the opportunity to provide insights into their software development approaches or techniques.

In this article, we seek to provide insights into our development by focusing on a) selected techniques that are relevant for developing more complex QGIS plug-ins, and b) the implementations of features that we found particularly challenging and not unique to the development of GIS4WRF. We address both, experienced developers with a background in developing Python packages or applications, and less experienced ones, whose work may focus on writing smaller data analysis and processing scripts and reusing existing packages.

The general structure of GIS4WRF and an overview of its two main components (*core* and *plugin*) and direct dependencies are given in Fig. 5. The *core* component is independent of QGIS and provides a collection of unit-tested (meta)data processing modules that use Python packages like NumPy (Oliphant, 2006) and netcdf4-python (Unidata, 2018). *System packages* are those that are available on any QGIS installation, while *extra packages* are installed using Python's package installation tool *pip* (PyPA, 2018) in a custom, user-level folder via the *bootstrap* component. The *plugin* component contains all the GUI source code and exposes the QGIS plug-in entry point. It relies on functionality from the *core* component and invokes the QGIS API.

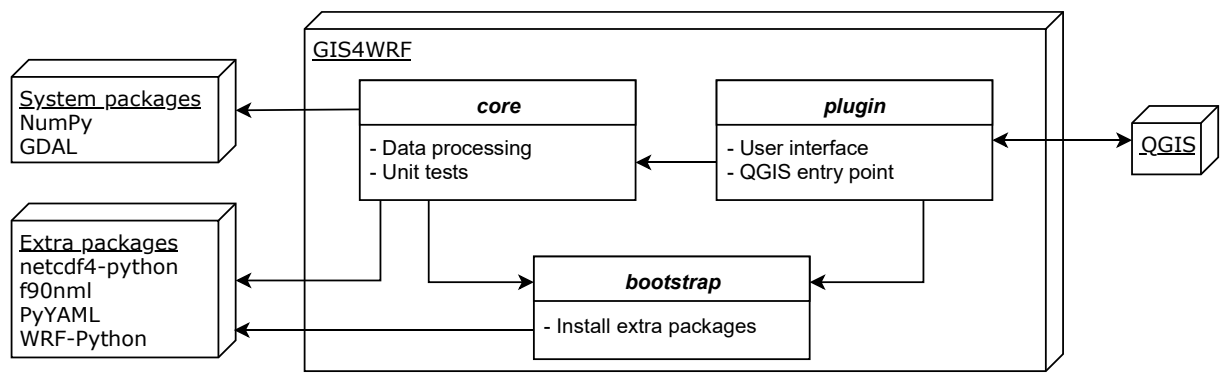

*Fig. 5. GIS4WRF architecture overview.*

## 4.1 External Dependencies

Python has a large ecosystem of scientific and engineering libraries (Python Software Foundation, 2018a) such as NumPy and SciPy (Jones et al., 2001). Developers can install such libraries, commonly referred to as packages, with package managers like *pip* or *conda* (Anaconda Inc., 2018).

When creating a new package in Python, developers need to list all its external packages, also called dependencies, that are required to run the source code of the new package. This is done in metadata contained in a file called *setup.py* (Python Software Foundation, 2018b). A QGIS plug-in is a Python package that, by design, cannot declare its dependencies. This means that, by default, a plug-in can only use the packages that are already installed within the Python installation that QGIS uses. For example, the system-wide installation on Linux and macOS, or a separate bundled installation on Windows.

If plug-in developers wish to use functionalities from a package that is normally not available in the QGIS Python environment, they can choose from the following common but, in our opinion, suboptimal, solutions: instruct the user for each supported platform on how to install the required package manually (e.g. QSWAT, UMEP), bundle the source code of the package within the plug-in, or re-implement the functionality. The first solution relies entirely on the users' ability to rectify problems during the installation and may lead to support enquiries to developers or system administrators; for example, a default QGIS installation on Windows requires administrator privileges to install new packages. For the same reason, trying to automate this approach is non-trivial. The second solution may be problematic for several reasons: the license of the package may be incompatible in terms of redistribution with the license of the plug-in, the package may contain binary extensions which must be compiled and bundled for each platform, or the package may use absolute module imports. The latter is an issue as the package modules would then reside within the plug-in package hierarchy and therefore require to be adapted.

In GIS4WRF we explored a different approach. At each time the plug-in is loaded, the *bootstrap* component checks for missing packages and installs them via Python's standard package manager *pip* into a platform- and plug-in-specific user-level folder that is subsequently added to the Python module search path. Updating the search path is required so that import statements can locate packages within this user-level folder. In our view, there are several advantages to this approach: the installation is automatic, no administrator privileges are required, and the installed packages do not affect other Python environments on the system. The trade-off lies in the way version constraints are handled. That is, if a plug-in requires a minimum version of a package, but the Python installation used by QGIS already contains a lower version of the same package, it is not possible to install and use the newer package version since Python only allows a single version per runtime environment. In this case, the plug-in notifies the user to manually update the package in question. Less common, as APIs of popular packages in Python are generally backwards-compatible, the same is true when requiring a maximum version. Similarly, if multiple enabled plug-ins use this approach and define the same dependencies, the plug-in that is loaded first has precedence over the others[4].

While our suggested approach combines many advantages and will only be problematic in corner cases, we argue that a better solution would be for QGIS to use a *virtual environment* (PyPA, 2016) per user, and, for each plug-in, to declare its dependencies. This would allow packages to be installed in a central location as part of a plug-in installation, including checks for version constraints for all installed plugins.

---

[4] Further technical details and the implementation of this component are available in the file *gis4wrf/bootstrap.py*.

## 4.2 Standard Packaging Conventions

A QGIS plug-in is a regular importable Python package, with the additional requirement that a pre-defined entry point must be provided for QGIS.

There are several ways to physically structure packages. In GIS4WRF, we follow the common practice recommended by the official Python Packaging User Guide (Python Software Foundation, 2018c), where the top-level folder of a package-type project contains ancillary files like *LICENSE.txt* and *README.md*, build scripts, and a subfolder for the package itself, in our case, *gis4wrf*. This layout provides a clear separation between the actual package and other supporting files.

We note that the majority of QGIS plug-in projects, likely unknowingly, break with this convention, and store the package contents directly in the top-level folder. By doing so, the creation of installable archives that only contain the package files itself, often requires to develop elaborate scripts where files are listed individually. Furthermore, to avoid having to add the parent folder of the top-level folder to Python's module search path, test modules must import plug-in modules by skipping one hierarchical level. This leads to different import statements between module types, e.g. *import mydialog* vs. import *myplugin.mydialog*, and pollutes the global namespace, potentially causing conflicts with external modules. We believe that this alternative convention has been observed so widely because many developers rely on the popular *Plugin Builder* tool for the generation of plug-in templates, and folder structure. In our view, especially when there is a need to develop complex plug-ins, developers should follow the recommendations highlighted above as per official Python Packaging User Guide.

## 4.3 Rapid Testing

Testing software, whether manually or automatically, plays an integral role in identifying mistakes. Our strategy for testing GIS4WRF involves two requirements such that a) non-GUI source code can be tested automatically outside QGIS, and using regular test frameworks, and b) any changes, whether GUI-related or not, are manually testable within a running instance of QGIS.

As described earlier, we split the plug-in into two layers, *core* and *plugin*, where *core* corresponds to non-GUI source code. We write tests for the *core* layer using the *pytest* (Krekel, 2018) framework. Similar to other test frameworks, *pytest* allows users to execute tests automatically from a command-line interface, giving an overview of successful and failed tests.

Manual testing, without restarting QGIS after each change means being able to reload the latest source code of a plug-in inside QGIS. The first challenge towards reaching that goal involves updating the installed plug-in package itself. Here, we choose to use a symbolic link to avoid the step entirely – we link the plug-in package folder to the location where QGIS stores installed plug-ins. Note that this is possible on UNIX-like systems, as well as on Windows. The second challenge is that once a module file has been imported, it is cached by the Python interpreter and not reloaded from disk on subsequent imports. This performance optimization means that updates to code are not automatically visible in a running Python instance. However, Python has APIs to force a reload of modules. This functionality can be invoked in QGIS by disabling and enabling the plug-in via the plug-in manager. Instead, we use the *Plugin Reloader* plug-in (Jurgiel, 2018) which exposes this functionality as a button within one of the toolbars of QGIS. These two techniques allow us to start manual testing of changed source code with a single mouse click.

### 4.4 Efficient Visualization

A major feature of GIS4WRF is the visualization of WRF-related data. Whether these are geographical input data in WPS binary format, or model output data in WRF-NetCDF format, our goals were to load and render data as fast as possible, while reusing any existing data reading, and rendering infrastructures, to reduce our development time.

Raster support in QGIS is provided by GDAL. While GDAL supports the NetCDF format (Rew and Davis, 1990), it relies on the CF Metadata Conventions to georeference the contained raster data. WRF-NetCDF files follow these conventions only partially, and omit, amongst others, standardized projection metadata. This leads QGIS to display the raster data in the wrong location and scale.

The technology we chose for reading and visualizing WPS binary and WRF-NetCDF data is GDAL's Virtual Datasets (VRT). VRTs are XML documents that describe how data on disk, or remote storage, are loaded and interpreted. VRTs load data through existing format drivers in GDAL, but also support reading raw numeric binary files. We use the former for reading WRF-NetCDF files, and the latter for WPS binary files. VRTs are handled like any other data format supported by GDAL. The power of a VRT comes from its ability to correct and fill missing metadata, e.g. spatial reference systems, without having to create a physical copy of the original dataset. They can be used to save pre-processing time and storage space. A VRT can also be assembled from several other raster datasets, including VRTs themselves, a useful feature required for the WPS binary format where data is sometimes split into multiple files, or tiles. An alternative to VRTs is a bespoke implementation of a *plugin layer* type in QGIS. A plugin layer implements how a dataset is read and converted to individual pixels. This approach, as used by the Crayfish plug-in (Lutra Consulting, 2018), allows it to be independent of GDAL, thus supporting more specialized data formats and advanced visualizations, including animations. A major disadvantage, however, is that QGIS GDAL-based raster-related processing tools cannot be applied to such layers. The Crayfish plug-in works around this problem by allowing to export the plugin layer as a GeoTIFF file which can then be opened again in QGIS without using the plug-in. We argue that, in the case of GIS4WRF, VRTs provide a better user experience while reducing overall development time.

# 5 Example Applications

For the purpose of demonstration, we show how GIS4WRF can be employed to create (high-resolution) third-party, land cover (LC) datasets and digital terrain models (DTMs) for use in WRF workflows.

Third-party derived LC datasets, that is LC datasets that are not included in the standard WPS datasets, have been employed in several studies (Brousse et al., 2016; Deng et al., 2015; Giannaros et al., 2018; Grossman-Clarke et al., 2010; Liao et al., 2014; Sertel et al., 2010). The MODIS datasets included in the standard WPS datasets are released with a global coverage and spatial resolution of 30 arcseconds and do not include yearly snapshots. Here, we process the 2016 yearly MODIS LC snapshot (Friedl and Sulla-Menashe, 2015) for the Conterminous United States (CONUS). The dataset is downloaded from NASA's Earthdata platform (https://earthdata.nasa.gov), merged, re-projected, and subsetted in QGIS for the CONUS area, and converted to the WPS binary format in GIS4WRF. The visualization of the result is done in QGIS with the help of GIS4WRF (Fig. 6).

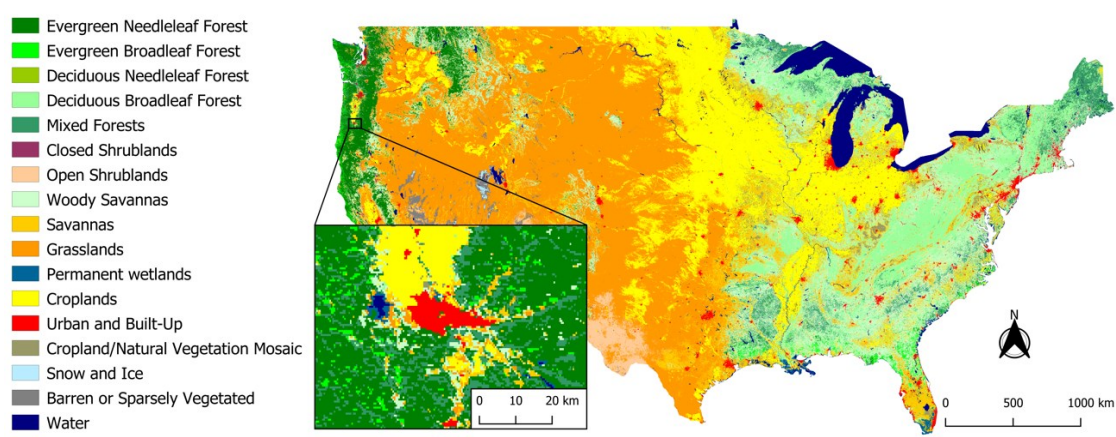

*Fig. 6. MODIS 2016 LC WPS binary format dataset for CONUS. Visualization in QGIS Print Composer.*

The use of high resolution DTMs in WRF has been employed in several studies (Göndöcs et al., 2017; Monaghan et al., 2013; Talbot et al., 2012). The Shuttle Radar Topography Mission (SRTM; Farr and Kobrick, 2000; van Zyl, 2001) was a collaboration between NASA's Jet Propulsion Laboratory and the National Geospatial-Intelligence Agency to produce a globally consistent digital elevation model for latitudes lower than 60° (Rodríguez et al., 2006). Global data are available for free at a spatial resolution of 1-3 arcseconds and can be downloaded from the USGS's (U. S. Geological Survey) EarthExplorer platform (https://earthexplorer.usgs.gov). With 10 to 30 times the horizontal spatial resolution of standard WPS datasets, the SRTM dataset can be used to improve the initialization of elevation data in high-resolution simulations. Here, we process the SRTM Version 3.0, 1 arcsecond void-filled data (U.S. Geological Survey (USGS), 2000) for Princeton, New Jersey, USA as used by Talbot et al. (2012) in their evaluation of real-case nested mesoscale large-eddy simulations with WRF. We begin with the creation of the innermost domain from the *Domains* subtab in GIS4WRF. We export the domain layer as a *GeoJSON* (Butler et al., 2016) file and use it to specify the download extent in EarthExplorer. After download, tiles are merged in QGIS, converted to WPS binary format and visualized (Fig. 7) using GIS4WRF.

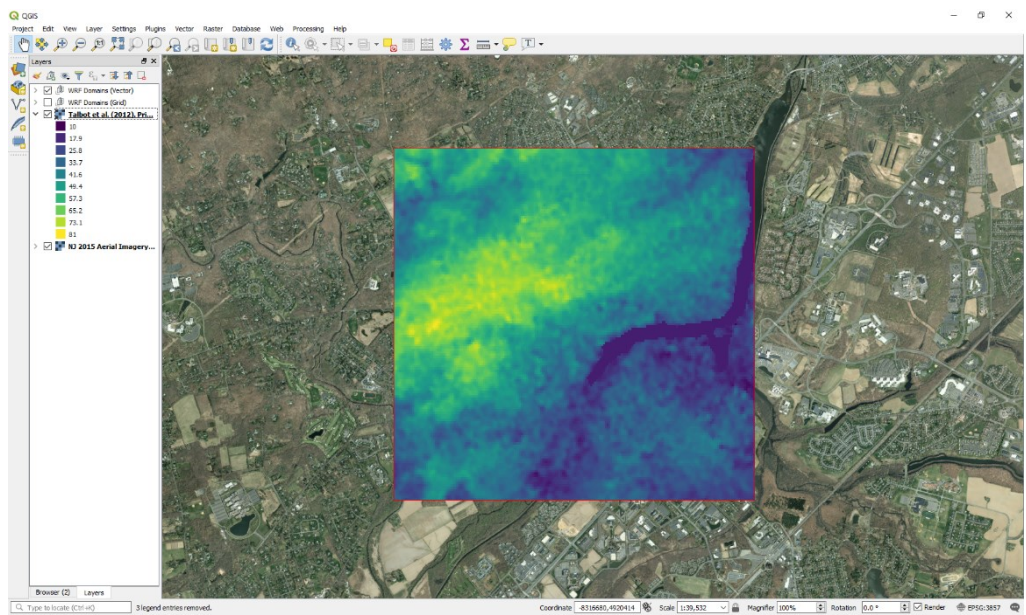

*Fig. 7. SRTM WPS binary format domain for Princeton, New Jersey, USA as used by Talbot et al.* (2012) *overlaid on* NJOIT and OGIS, (2015) *base map.*

# 6   Conclusions and Future Work

We presented GIS4WRF, a cross-platform QGIS plug-in written in Python that incorporates new and existing WRF tools for data processing, configuration, simulation, and visualization into a single graphical environment. GIS4WRF enriches the capabilities of WRF tools with a large community-based and growing ecosystem of geoscientific software. The key contributions of GIS4WRF are evident in the way it streamlines WPS and WRF workflows, and in its ability to simplify WRF-related tasks, thus allowing for a greater number of users to access advanced features and save time on tedious and time-consuming tasks.

We discussed the main challenges during its development, showed the limitations of current approaches, and described possible avenues to overcome them. These include the handling of external Python package dependencies through a *bootstrap* component, enabling automated software testing by factoring into architectural layers, and using GDAL's Virtual Dataset mechanism for efficient visualization of non-standard data formats.

GIS4WRF is developed as a free, open source and community-driven project on GitHub at https://github.com/GIS4WRF. We strongly encourage users to provide feedback, bug reports and feature requests, via GitHub's issue system at https://github.com/GIS4WRF/gis4wrf/issues. Depending on the feedback and general uptake by the community, future GIS4WRF releases may include WUDAPT (Ching et al., 2018) support, integration with cloud providers, visualization of GRIB data, integration with multiple providers of meteorological data (e.g. ECMWF Web API), and additional dataset conversion tools.

## Software Availability

| **Software Name** | GIS4WRF |
|---|---|
| **Developers** | D. Meyer and M. Riechert |
| **Programming Language** | Python :: 3.6 |
| **Software License** | MIT |
| **Cost** | Free |
| **Year First Available** | 2018 |
| **Software Required** | QGIS 3 |
| **Availability (in QGIS)** | From *Plugins > Manage and Install Plugins …* |
| **Website** | https://gis4wrf.github.io |
| **Manual** | https://gis4wrf.github.io/documentation |
| **Tutorials** | https://gis4wrf.github.io/tutorials |
| **Code repository** | https://github.com/GIS4WRF/gis4wrf |
| **Contact** | https://github.com/GIS4WRF/gis4wrf/issues |

# Appendix – Domain Calculations

In WRF, a nested domain is always defined relative to its parent domain. In GIS4WRF, a nested domain is always defined relative to its innermost domain, thus allowing for a greater control over the location of the domain of interest (i.e. the innermost domain).

The method we developed to convert from the WRF domain reference system to the GIS4WRF domain reference system is summarized below. Although the method is general, for simplicity, we show a case formed of 2 domains: an innermost- and an outermost-domain referred to as child- and parent domain, respectively (Fig. A1). A list of symbols, and corresponding variable names used in WRF and names used in the GIS4WRF interface are summarized in table A1.

*Table A1*
*List of Symbols used in appendix and corresponding names used in the WRF User Guide* (Wang et al., 2017)*, and in the GIS4WRF interface respectively.*

| Symbol | WRF Variable | GIS4WRF Name | Unit | Description |
| --- | --- | --- | --- | --- |
| $x_{child}^{min}$ | n/a | n/a (computed) | °, m | Child domain minimum x coordinate |
| $y_{child}^{min}$ | n/a | n/a (computed) | °, m | Child domain minimum y coordinate |
| $x_{child}^{max}$ | n/a | n/a (computed) | °, m | Child domain maximum x coordinate |
| $y_{child}^{max}$ | n/a | n/a (computed) | °, m | Child domain maximum y coordinate |
| $x_{child}^{centre}$ | n/a (computed) | Center Point Longitude | °, m | Child domain half-width coordinate |
| $y_{child}^{centre}$ | n/a (computed) | Center Point Latitude | °, m | Child domain half-height coordinate |
| $\Delta L_{child}$ | n/a (computed) | Horizontal Grid Spacing | °, m | Child domain horizontal grid spacing (resolution) |
| $cols_{child}$ | E_WE | West to East Grid Extent | - | Child domain total number of columns (extent) |
| $rows_{child}$ | E_SN | South to North Grid Extent | - | Child domain total number of rows (extent) |
| $x_{parent}^{min}$ | n/a | n/a (computed) | °, m | parent domain left corner |
| $y_{parent}^{min}$ | n/a | n/a (computed) | °, m | parent domain bottom corner |
| $x_{parent}^{max}$ | n/a | n/a (computed) | °, m | parent domain right corner |
| $y_{parent}^{max}$ | n/a | n/a (computed) | °, m | parent domain top corner |
| $x_{parent}^{centre}$ | KNOWN_LON | n/a (computed) | °, m | parent domain half-width coordinate |
| $y_{parent}^{centre}$ | KNOWN_LAT | n/a (computed) | °, m | parent domain half-height coordinate |
| $\Delta L_{parent}$ | DX | n/a (computed) | °, m | parent domain horizontal grid spacing (resolution) |
| $pad_{parent}^{left}$ | n/a | Left Padding | - | Number of columns to the left of child domain |
| $pad_{parent}^{bottom}$ | n/a | Bottom Padding | - | Number of rows to the bottom of child domain |
| $pad_{parent}^{right}$ | n/a | Right Padding | - | Number of columns to the right of child domain |
| $pad_{parent}^{top}$ | n/a | Top Padding | - | Number of rows to the top of child domain |
| CPR | PGR | Child-to-Parent Ratio | - | Ratio of child to parent domain horizontal grid spacing. |

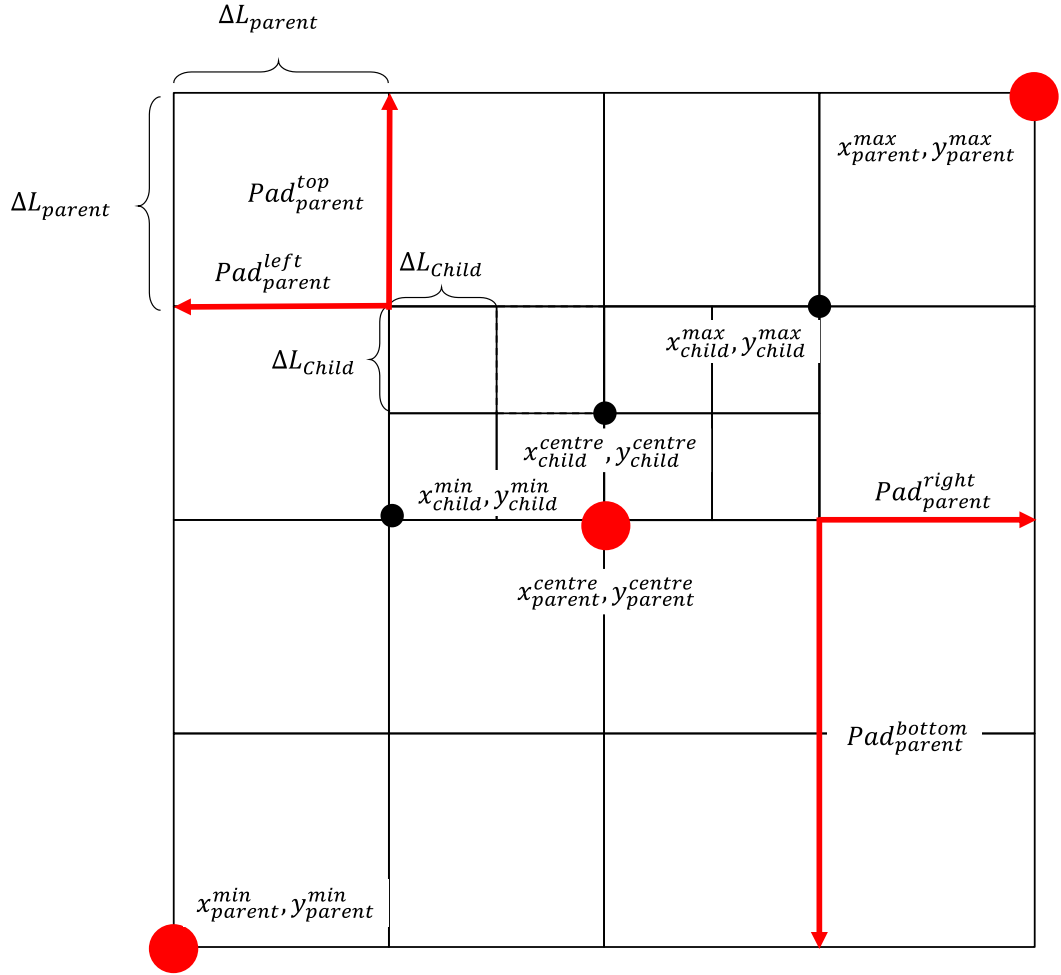

*Fig. A1. Idealized 4 x 2 domain and a 4 x 4 outer domain with child-to-parent horizontal grid spacing ratio of 2.*

The calculation of outer domains from inner domains consists of four steps followed by a check. First, we compute the child domain bottom left and top right corner coordinates from its center coordinates, horizontal grid spacing and number of grid cells:

$$x_{child}^{min} = x_{child}^{centre} - \frac{\Delta L_{child} \times cols_{child}}{2},$$
$$y_{child}^{min} = y_{child}^{centre} - \frac{\Delta L_{child} \times rows_{child}}{2},$$
$$x_{child}^{max} = x_{child}^{centre} + \frac{\Delta L_{child} \times cols_{child}}{2},$$
$$y_{child}^{max} = y_{child}^{centre} + \frac{\Delta L_{child} \times rows_{child}}{2}.$$

Second, we compute the parent domain horizontal grid spacing from the child to parent horizontal grid ratio and the child domain horizontal grid spacing:

$$\Delta L_{parent} = \text{CPR} \times \Delta L_{child}.$$

Third, we compute the parent domain bottom left and top right corner coordinates from the child domain four corner coordinates, parent domain horizontal grid spacing and parent domain padding cells:

$$x_{parent}^{min} = x_{child}^{min} - \frac{\Delta L_{parent} \times pad_{parent}^{left}}{2},$$
$$y_{parent}^{min} = y_{child}^{min} - \frac{\Delta L_{parent} \times pad_{parent}^{bottom}}{2},$$
$$x_{parent}^{max} = x_{child}^{max} + \frac{\Delta L_{parent} \times pad_{parent}^{right}}{2},$$
$$y_{parent}^{max} = y_{child}^{max} + \frac{\Delta L_{parent} \times pad_{parent}^{top}}{2}.$$

Fourth, we compute the parent domain center coordinates from its bottom left and top right corner coordinates:

$$x_{parent}^{centre} = \frac{x_{parent}^{min} + x_{parent}^{max}}{2},$$
$$y_{parent}^{centre} = \frac{y_{parent}^{min} + y_{parent}^{max}}{2}.$$

WRF domains require the number of columns and rows in the child domain to be an integer multiple of the nest's parent domain (Wang et al., 2017). When we calculate the parent domain, we check that the number of columns and rows of each child domain divided by their respective child to parent horizontal grid spacing ratio ($CPR$) is an integer; else we add one grid cell to the child domain columns or rows until we satisfy the condition.